\documentclass[journal=jacsat,manuscript=article]{achemso}
\usepackage[version=3]{mhchem} 

\usepackage{graphicx,subfigure}
\usepackage{dcolumn}
\usepackage{bm}
\usepackage{epsfig}
\usepackage{latexsym}
\usepackage{amsfonts}
\usepackage{amssymb}
\usepackage{mathrsfs}
\usepackage{subfigure}
\usepackage{setspace}

\title{ M13-phage-based star-shaped
particles with internal flexibility\\
}

\author{Arantza B. Zavala-Martínez}
\author{Eric Grelet}%
\email{eric.grelet@crpp.cnrs.fr}

\affiliation{%
Univ. Bordeaux, CNRS, Centre de Recherche Paul-Pascal, UMR 5031,\\ 115 Avenue du Dr. Schweitzer, F-33600 Pessac, France\\}%

\begin{document}

\begin{abstract}

We report on the construction and the 
dynamics of monodisperse star-shaped particles, mimicking, at the mesoscale, star polymers.
Such multi-arm star-like particles result from the self-assembly of gold 
nanoparticles, 
forming the core, with tip-linked 
filamentous viruses -- M13 bacteriophages -- acting as spines in a sea urchin-like structure. 
By combining fluorescence and dark-field microscopy with dynamic light scattering, we investigate the 
diffusion of these hybrid spiny particles. 
We reveal the internal dynamics of the star particles
by probing their central metallic core, which exhibits a hindered motion 
that can be described as a Brownian particle trapped in a harmonic potential. We therefore show that the filamentous viruses and specifically their tip proteins behave as entropic springs, 
extending 
the relevance of the study of such hybrid mesoscopic analogs of star polymers 
to phage biotechnology.
\end{abstract}


\clearpage

So-called ``soft particles" refer to a class of objects with a dual character somewhere between hard colloids
and polymer coils \cite{vincent_depletion_1986,Likos2006,VLASSOPOULOS2014561}.
Common examples of these particles include microgels \cite{wei_mechanism_2013, lindenblatt_synthesis_2000,Eckert2008}, micelles \cite{laurati_starlike_2005, merlet-lacroix_swelling_2010} and star polymers \cite{erwin_dynamics_2010,Gupta2015,gury2019colloidal}, whose phase behaviors and dynamics change as the level of their softness increases. 
For hard spherical particles, which undergo transitions from a disordered fluid phase to a 
crystalline phase at relatively low volume fractions \cite{pusey_phase_1986}, their Brownian diffusion can be affected by hydrodynamic or inter-particle interactions \cite{tokuyama1994dynamics, pusey1983hydrodynamic, van1991dynamics}. 
In contrast, ultra-soft particles such as polymer coils behave as amorphous fluids even at very high volume fractions \cite{vlassopoulos2021suspensions}, and their dynamics is controlled by the elasticity of the chains, the deformability of the core, and entanglement \cite{de1982dynamics, chu1991dynamic}.
For systems in between, including microgels, made of cross-linked polymeric networks swollen by a solvent, or star polymers, consisting of long polymeric arms connected to a central core, the dynamical behavior depends also on the compressibility and interdigitation of the particles \cite{romeo2012origin, freedman2005diffusion,Cautela2020}.
These interactions for soft systems can suppress crystallization and lead to direct transitions to glassy and jammed states
\cite{gasser_form_2014, wei_mechanism_2013, laurati2005starlike, fleischer_self-di_2000}.
Thus, the design of custom-made particles 
intermediate between the hard and ultra-soft limits plays a key-role for the general understanding of phenomena 
such as 
the origin of glass transition \cite{Mattsson2009} or the formation of densely packed states \cite{Conley2017,Manolis2014,Manolis2022}.

 
\begin{figure*}[htb]
\includegraphics[width=0.75\linewidth]{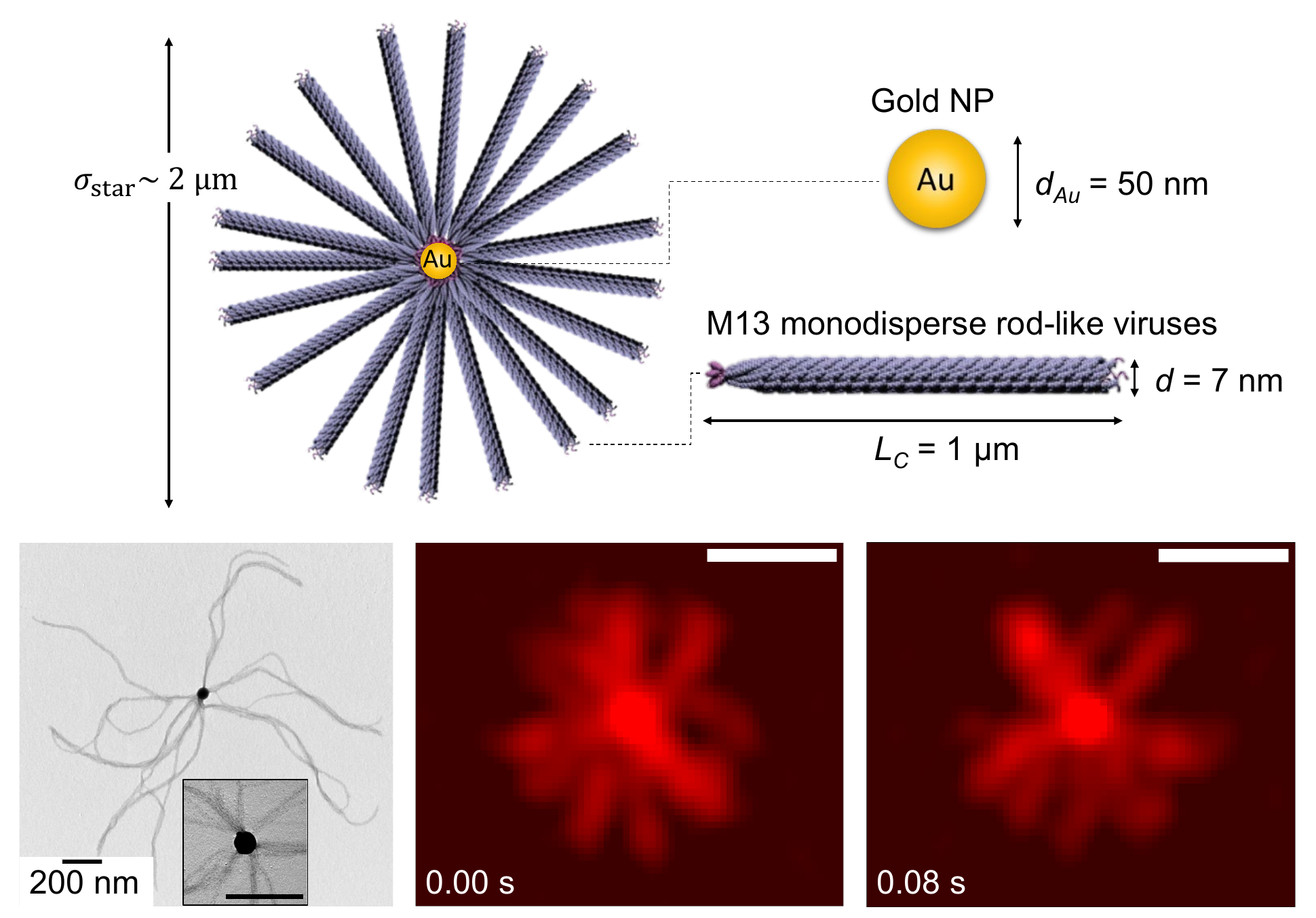}
\caption{\textbf{Hybrid M13-virus-based star particles. } 
Top: Schematic representation of the particle structure exhibiting an urchin-like morphology and its two components: A gold nanoparticle (Au NP) as core, and arms or spines formed by tip-linked monodisperse semi-flexible rod-like viruses (M13 phage). Bottom: Transmission electron microscopy image (left) and confocal fluorescence microscopy images (right) of star particles. The white scale bars 
represent $\mathrm{1 \ \mu m}$.  }
\label{fig:Star-components} 
\end{figure*}

In this context, multi-arm star-like systems, where the number, size and flexibility of the arms are tunable, leading to variable deformability and interdigitation provide an ideal testbed for the study of the impact of ``softness'' on dynamics
\cite{buitenhuis_block_1997, vlassopoulos_multiarm_2001}. 
Polymer chains grafted on a nanosphere core represent the most simplified star-like system with tunable softness.
In the limit of chains short compared to core size, the hard sphere behavior may prevail, whereas for long polymer chains, the structure and the dynamics of ultra-soft particles are expected \cite{cloitre_high_2010, choi_toughening_2012}. 
For instance, star-like particles such as polymer-grafted latex or silica nanoparticles in the range of sizes from tens to hundreds of nanometers have already been achieved 
 and used to investigate glass formation by varying the strength of interparticle interactions \cite{shay2001thermoreversible, ohno2006suspensions, ohno_suspensions_2007}. 
However, the main limitation of these polymer-grafted nanospheres is their high polydispersity in both size and number of arms. 
Moreover, the polymer chains are in the high flexibility limit, where their persistence length, $L_p$, is short compared to their contour length $L_c$ \cite{tricot1986chain, lee_molecular_2008}, \textit{i.e.} $L_p \ll L_c$. 

Recently, DNA nanotechnology has provided an exquisite tool for the construction of monodisperse systems of multiarm nanostar particles \cite{biffi_phase_2013, Rovigatti-Sciortino, brady_crystallization_2017, Lattuada-Sciortino}.
Monodisperse in size, these DNA molecular stars provide a new model system for soft matter and nanoscience, including soft particle dynamics. 
They differ from the polymer-grafted nanospheres in having a low number of very stiff arms, with
a persistence length of 
$L_p \simeq 50 \ \mathrm{nm}$  \cite{lu_dna_2002}, and an arm contour length, $L_c$, of a few nanometers \cite{brady_flexibility_2019, biffi_phase_2013}, putting them in the $L_c \ll L_p$ limit. 
They are, however, difficult to study at the single particle level because of their 
limited size. At higher length scale, colloids coated with ultralong DNA fragments resulting in dense spherical charged brushes, have been designed and their phase behavior reveals fundamental properties in terms of deformation and interpenetration of the particles.\cite{Manolis2014,Manolis2022}


Here, we propose a model system of monodisperse soft star-like particles 
composed of a gold nanoparticle core -- Au NP --, linked with semi-rigid filamentous viruses -- M13 bacteriophages -- acting as spines in an urchin-like structure \cite{delaCotte2017,Zhan2022}, 
as shown in Fig.~\ref{fig:Star-components}.
Thanks to their biological origin, these rod-shaped viruses are uniform in size (See Materials and Methods). They have been genetically modified to regio-selectively interact \textit{via} one of their tips with Au, \cite{Mosquera2020} and have been already used in sensor application.\cite{Zhan2022} The self-assembly between the metallic (Au NPs) and the biological (M13 phages) components results in sparse particles which are monodisperse in size and whose number of arms can be tuned.  
The persistence length of these viral arms is comparable to their contour length, \textit{i.e.}, $L_p \sim L_c$ \cite{song1991dynamic},  
putting them in between the polymer-grafted nanospheres and the DNA nanostars. 
The relative dimension of the Au NP core to the overall micrometer length of the virus arms 
makes our hybrid particles a scaled-up system of soft star-like particles, whose components can be directly tracked by optical microscopy.
In this paper, we report on the dynamics of an ensemble of virus-based star particles in the dilute regime --- where the inter-particle interactions are negligible ---,  
by means of single particle tracking (SPT) and dynamic light scattering (DLS).
Our results show diffusive behavior on long time scales but are also able to resolve the change of dynamics on short time scales due to the internal degrees of freedom of our particles.

\section*{Results}

We construct our star-like particles \textit{via} the self-assembly in aqueous solution of 50~nm Au nanoparticles (Au NPs) with micrometer long filamentous rod-like viruses (M13) which are genetically modified to selectively and irreversibly bind to noble metals with one of their tips (see Materials and methods). This results in a hybrid particle, whose Au core is surrounded by a brush formed by about $n \simeq 23$ semi-flexible viral arms (Fig.~\ref{fig:Star-components} and Supplementary Fig.~S1). The rigidity of the viral arms whose persistence length is larger than their contour length ($L_p>L_c$, see Materials and Methods) imparts a 
sea urchin-like morphology to the particles. 
A key-feature of these isometric particles is their \textit{monodispersity} in diameter, $\sigma_\mathrm{star}\simeq 2~\mu$m (Fig.~\ref{fig:Star-components}) stemming from the intrinsic uniformity in size of the viral ``spines" \cite{delaCotte2017}. 

\begin{figure} 
\subfigure[]{\includegraphics[ 
width=0.65\linewidth]{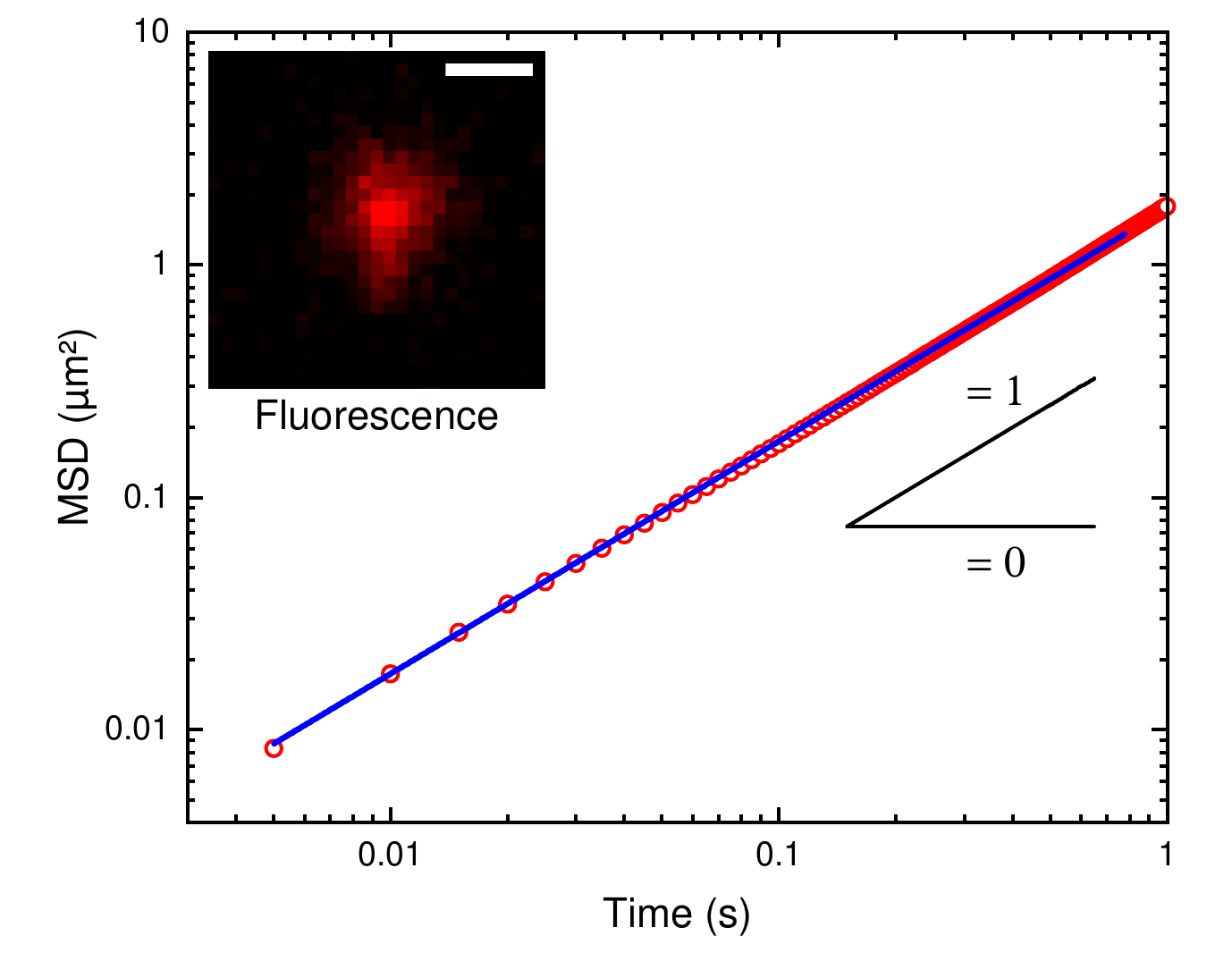}}
\subfigure[]{\includegraphics[
width=0.65\linewidth]{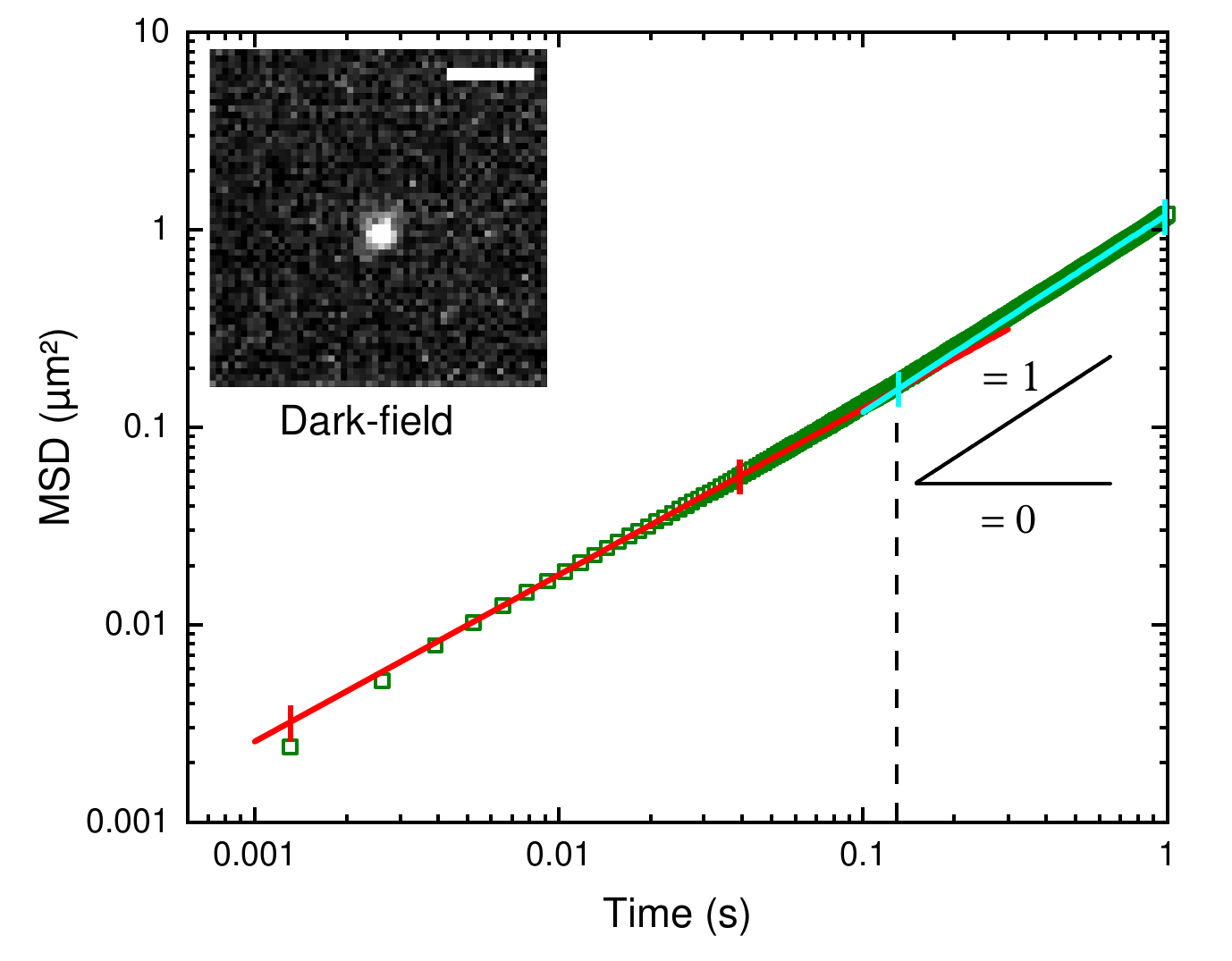}}
\caption{\textbf{Star particle mean squared displacement (MSD) measured by (a) fluorescence and (b) dark-field microscopy}. 
(a) The blue line is a power-law fit according to Eq.~\ref{MSD-eqfit} and indicates a diffusive regime ($\gamma = 1$) over the full range of the times probed, with a diffusion coefficient of $D_{T}^{Fluo} = 0.43~\mathrm{\mu m^2/s}$. Inset: Fluorescence microscopy image of a star particle obtained during single particle tracking performed for the MSD determination. 
(b) Dark-field microscopy experiment, where solely the Au core of the star is tracked as shown in the inset. 
The lines are power-law fits according to Eq. \ref{MSD-eqfit}, and indicate two regimes: a diffusive behavior (cyan line, $\gamma = 1$) at long times with a diffusion coefficient $D_{T}^{DF} = 0.30~\mathrm{\mu m^2/s}$, and a subdiffusive one (red line) with $\gamma = 0.85$ for short times.   
A crossover time $\tau_{cross} \simeq$~130~ms can be defined between the two regimes, and it is shown by the vertical dashed line. The fit range of each regime is indicated by small colored vertical dashes. In both insets, the scale bar represents $2~\mathrm{\mu m}$.}
\label{fig:MSD}
\end{figure}

The confocal microscopy images of Fig.~\ref{fig:Star-components} and the Supplementary Movies show that, beyond the Brownian displacement of the particles, their viral arms are also moving. For a better understanding of the star behavior, their dynamics is investigated in the dilute regime by single particle tracking using fluorescence microscopy. Traces are recorded in the observation plane (Supplementary Fig.~S2), \textit{i.e.}, nearly in 2D,
from which the mean squared displacement (MSD) is calculated (see Materials and Methods for details and Supplementary Fig.~S3), as plotted in Fig.~\ref{fig:MSD}(a). 
The general expression of the mean squared displacement in $n$ dimensions as a function of time is \cite{Crocker1996,Eric-Laura}: 
\begin{equation}
\label{MSD-eqfit}
\mathrm{MSD} = 2 n D_T t^\gamma \mathrm{,}
\end{equation}

\noindent with $D_T$ the translational diffusion rate and $\gamma$ the diffusivity exponent indicating the nature of the dynamics: subdiffusive for $\gamma<1$, diffusive for $\gamma = 1$, and superdiffusive, for $\gamma>1$. 
The linear dependence with time, found for the 
MSD measured by fluorescence (Fig.~\ref{fig:MSD}(a)), indicates a diffusive regime ($\gamma = 1$) for the star particles, from which the three-dimensional translational diffusion coefficient $D_{T}^{Fluo}= 0.43~\mathrm{\mu m^2/s}$ can be determined according to Eq.~\ref{MSD-eqfit} with $n=2$. 

However, a careful inspection of the tracking by fluorescence microscopy suggests that the measured MSD and therefore $D_{T}^{Fluo}$ are overestimated because they include a random uncertainty in the determination of the particle center-of-mass, stemming from the motion of the arms, which shifts the maximum fluorescence intensity position from the real particle center-of-mass (see Supporting Information, Inset of Fig.~\ref{fig:MSD}(a) and Supplementary Fig.~S4). 

To circumvent this problem, we take advantage of the hybrid feature of the stars, and specifically that the Au core of the particle exhibits a high scattering signal as observed by dark-field (DF) microscopy. Using this technique, we are able to visualize only the star's metallic core, as the viral organic arms are nearly invisible with this method (see inset of Fig.~\ref{fig:MSD}(b)). 
Single particle tracking is then performed and the high contrast of the Au core revealed by dark-field allows for a significant increase of the frame rate of the acquired movies (see Materials and Methods) giving us access to shorter time scales, \textit{i.e.} $\sim 1 \ \mathrm{ms}$. 
The resulting MSD averaged over about 160 traces is shown in Fig.~\ref{fig:MSD}(b) and Supplementary Fig.~S3. Interestingly, two regimes can be distinguished: a diffusive behavior ($\gamma = 1$) at long times from which a diffusion coefficient $D_{T}^{DF}=0.30 \ \mathrm{\mu m^2/s}$ is measured, and a subdiffusive regime with $\gamma \simeq 0.85 $. The former regime shows a diffusion coefficient, $D_{T}^{DF}$, significantly lower than that measured with fluorescence and confirms \textit{a posteriori} that  $D_{T}^{Fluo}$ is overestimated, whereas the subdiffusivity of the latter regime suggests some hindered motion of the metal core. 

For a better picture of what happens within the components of the star and to complete the investigation of the different MSD regimes found in Fig.~\ref{fig:MSD}(b), we complement the study by using dynamic light scattering (DLS) in the dilute regime of particle concentration. In DLS as in DF microscopy, the scattering signal of the Au NPs far exceeds that of the viruses.
The intensity correlation function, or second-order correlation function, $g_2(q,t)$, is acquired as a function of the scattering wavevector magnitude, $q$, which is directly related to
 the scattering angle $\theta$ (see Eq.~\ref{Scattering Angle} in Materials and Methods).\cite{Stetefeld2016} 
In the simplest case, \textit{i.e.} for non-interacting, solid, spherical particles uniform in diameter, 
$g_2(q,t)$ is a single exponential decay function with decay time, $\tau = (2q^2 D_T)^{-1}$, expressed in terms of the self-intermediate scattering function, $F_s(q,t)$, as \cite{Pecora}
		
\begin{equation}
\label{g2-spheres}
g_2(q,t) - 1 = \beta \exp{(-2q^2 D_T t)} = \beta |F_s (q,t)|^2 \mathrm{,}
\end{equation}

\noindent with $\beta$, the coherence factor, 
expected to be close to 1 for nearly ideal experimental conditions. 
The experimental intensity correlation functions $g_2(q,t)$ acquired by DLS are \textit{globally} fitted using Eq.~\ref{g2-spheres}, as shown in Supplementary Fig.~S5.  
A simple inspection of the global fit indicates that Eq.~\ref{g2-spheres} cannot properly account for the experimental scattering signal of the stars, showing a significant deviation of their dynamics from the standard Brownian motion of non-interacting uniform solid spheres \cite{Pecora}. 

This, in combination with the subdiffusive regime of the MSD observed at short times by DF microscopy (Fig.~\ref{fig:MSD}(b)), suggests that the grafting of viruses to the Au nanoparticle surface inhibits the motion of the Au core around its reference 
position at the center of the star particle. We propose, therefore, 
as a description of the internal dynamics of the stars, to model the hindered motion of their core by a Brownian particle trapped in a harmonic potential (spring model). 

\begin{figure}[htb]
\centering
\includegraphics[width=0.7\linewidth]{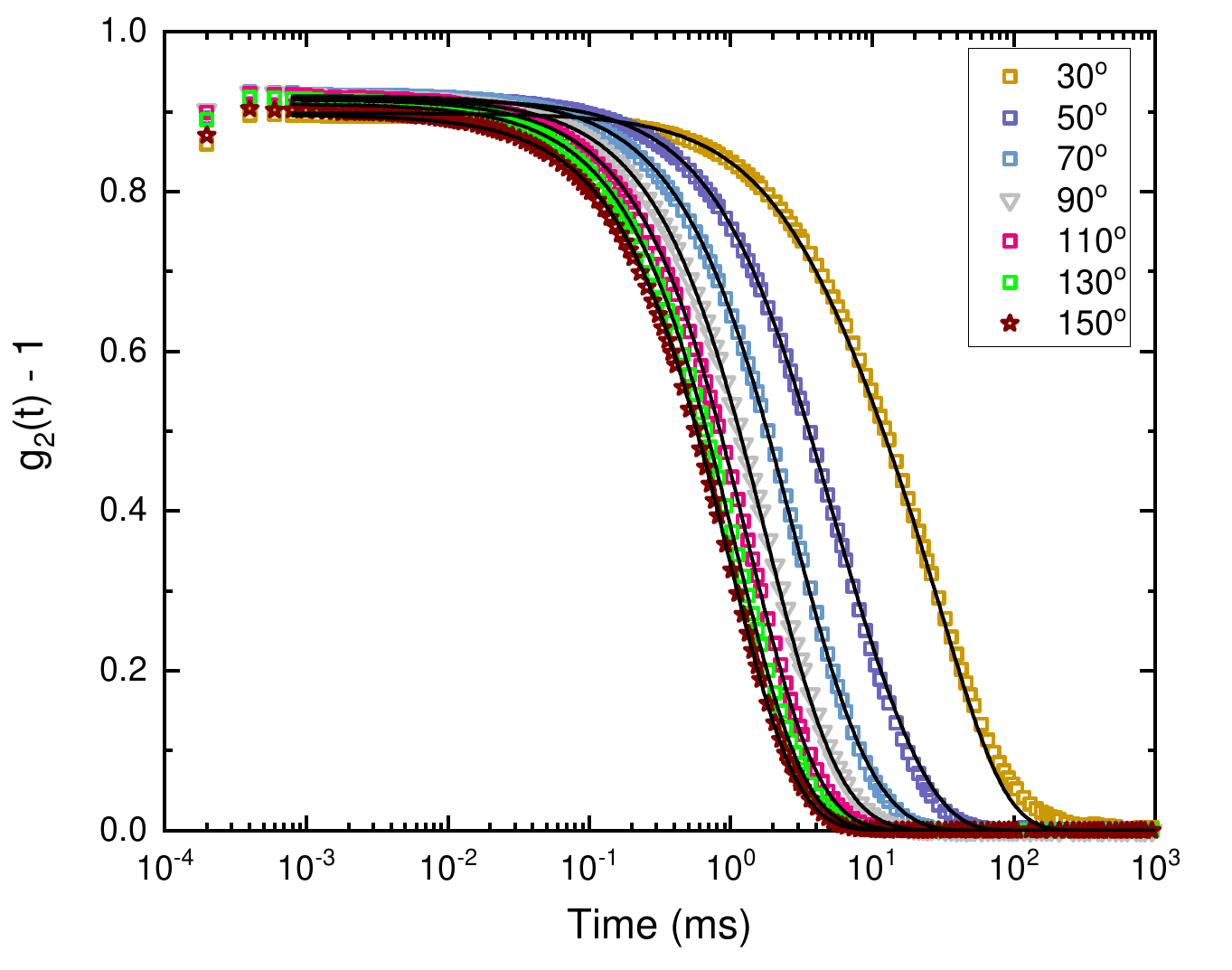}
\caption{\textbf{Intensity correlation function \textit{g$_2$}(\textit{q,t}) of star particle suspensions at $\mathbf{OD_{520}=0.15}$ measured by dynamic light scattering} for various scattering angles $\theta$ (symbols) from $30^\mathrm{o}$ to $150^\mathrm{o}$ in steps of $10^\mathrm{o}$. For clarity, only half of the angles, \textit{i.e.} every $20^\mathrm{o}$, have been plotted. Data are \textit{globally} fitted using the spring model (Eq.~\ref{fit-spring}) 
(black lines). The global parameters obtained from the fit are: $D_T = 0.33 \pm 0.02 \ \mathrm{\mu m^2/s}$, $\left\langle  r^2_{int} \right\rangle = 2.7 \pm 0.4 \times 10^{-3} \ \mathrm{\mu m^2}$ and $\xi = 0.17 \pm 0.01 \ \mathrm{ms^{-1}}$.}
\label{fig:DLS-CS}
\end{figure}

This process was 
first described  
by Ornstein \& Uhlenbeck \cite{Onrstein-Uhlenbeck}, and leads to an extra-
contribution, $F_{int}$, to the self-intermediate scattering function $F_s$ (see Materials and Methods) \cite{Fragments-link}, related to the intensity correlation function through Eq.~\ref{g2-spheres}


\begin{align}
\begin{split}
\label{fit-spring}
g_2(q,t) -1 =  \beta \lbrace   & \exp{(-D_T q^2 t)} \\
 &\times \left(  \exp  \left\lbrace   -q^2 \left\langle  r^2_{int} \right\rangle   \left[1 - \exp(-\xi t) \right] \right\rbrace\right)  \rbrace^2    \mathrm{,}
\end{split}
\end{align} 
	
\noindent with $\left\langle  r^2_{int} \right\rangle = \frac{k_B T}{\kappa}$ the mean square displacement within the harmonic potential at the relaxation time $1/\xi = \gamma_0/ \kappa$, written in terms of the friction coefficient, $\gamma_0$, and spring constant, $\kappa$, associated with the harmonic potential. 
The \textit{global} fit of the intensity correlation functions at different angles using the spring model (Eq.~\ref{fit-spring}) is represented by the black lines in Fig.~\ref{fig:DLS-CS}. It accounts accurately for the experimental scattering signal, with the following global parameters: $D_T = 0.33 \pm 0.02 \ \mathrm{\mu m^2/s}$, $\left\langle  r^2_{int} \right\rangle = 2.7 \pm 0.4 \times 10^{-3} \ \mathrm{\mu m^2}$ and $\xi = 0.17 \pm 0.01 \ \mathrm{ms^{-1}}$. In order to prove that these results are obtained in the dilute regime where the interactions between particles can be neglected, the same analysis is performed for other dilutions as shown in Supplementary Figs.~S6 and S7 for which the fit parameters (see caption of the Figs.~\ref{fig:DLS-CS}, S6, and S7) are found to be independent of the sample dilution.

\section*{Discussion}

The diffusion coefficient $D_T = 0.33 \ \mathrm{\mu m^2/s}$, obtained by DLS using Ornstein \& Uhlenbeck model is in outstanding agreement with the one determined by SPT, $D_{T}^{DF}=0.30 \ \mathrm{\mu m^2/s}$, considering the independence of the two approaches and the number of 
independent parameters (three, namely $D_{T}$, $\left\langle  r^2_{int} \right\rangle$ and $\xi$) used for fitting the DLS data. 


From that diffusion coefficient, the hydrodynamic radius, $R_H$, 
referring to the size of the hard sphere diffusing at the same $D_T$ can be defined as $R_H=\ k_B T / (6\pi\eta D_T)$ with $\eta$ the viscosity of water, and is found to be $R_H=0.65~\mu$m. Such a value is lower than the star geometric radius 
or equivalently to the virus contour length, \textit{i.e.} $R_H < \sigma_\mathrm{star}/2 \simeq L_c = 1~\mu$m, a result which is expected in view of the fuzzyness and loose morphology of the particles compared to a solid sphere.


Two other physical parameters can be extracted from the global fit (see previous Section and Fig.~\ref{fig:DLS-CS}) using Ornstein \& Uhlenbeck model: 1)~the mean square displacement $\left\langle  r^2_{int} \right\rangle = \frac{k_B T}{\kappa}$ within the harmonic potential,  
from which the spring constant can be directly deduced, $\kappa = 370 \ k_B T / \mathrm{\mu m^2}$; and 2)~the relaxation time $1/\xi = \gamma_0/ \kappa$, which allows for the determination of the friction coefficient, $\gamma_0 = 2.2 \ k_B T \ \mathrm{s}/ \mathrm{\mu m^2 }$. The Stokes radius, $R_S$, corresponding to the size of a spherical dense colloid having the same friction $\gamma_0 = 6\pi\eta R_S$ and which is therefore conceptually equivalent to a hydrodynamic radius, is found to be $R_S=0.48~\mu$m. This value derived from the second and third fitting parameters is 
smaller by about 26\%  than the hydrodynamic radius $R_H$ derived above from the first fitting parameter. 
While $R_H$ is associated with the motion of the entire star in aqueous solvent, $R_S$ is related to the internal dynamics of the particle and the displacement of the core in the crowded environment of viruses.

As a consequence, our modeling evidences that the star particles exhibit internal degrees of freedom, which can stem from the tip proteins p3 used for binding the viruses to the Au NPs and acting as entropic springs, as well as from the flexibility of the whole phages quantified by their persistence length $L_p$. In the latter case, the phage effective length or end-to-end length $L$ is a function of the contour length $L_c$ and the persistence length $L_p$, which leads to $L=0.94~\mu$m  
according to Kratky \& Porod relation \cite{Kratky1949b} for isolated polymer chains. The difference between the phage contour length and effective length $L_c-L \simeq 60~$nm is consistent with the typical length scale obtained from the root mean square displacement $\sqrt{r^2_{int}}\sim 52~$nm (Fig.~\ref{fig:DLS-CS}). However, visual inspection of the Supplementary Movie acquired by fluorescence microscopy reveals that the viral arms have a significant angular motion beyond the rotational diffusion of the full particle, which suggests a more localized flexibility associated with the tip proteins p3 linking the M13 virus 
to the Au NP. The flexibility of the p3 tip protein     
is consistent with its biological structure comprising 406 amino acids 
formed by three domains (N1, N2, and C) separated by soft flexible glycine-rich linkers \cite{Marvin2014}. These glycine-rich linkers result in the spreading of p3 away from the virus body and the presence of knobs as seen in electron micrographs \cite{gray1981adsorption}, and they provide to the protein its ability for conformational change. Moreover, the elastic and structural flexibility 
of p3 has been recently shown to play a major role in the infection of the bacterial host \cite{Conners2023}. In its standard configuration, as used here, the spatial extension of the five p3 tip proteins far exceeds the 7~nm of the virus diameter 
and has been found to be around $\ell \sim 20~$nm, enabling a high accessibility \cite{gray1981adsorption,Conners2023}. 
The corresponding footprint of the p3 proteins can then be estimated by $\pi (\ell/2)^2$ and be compared with the area available on the Au NP surface (neglecting curvature effect): $\pi d_{Au}^2$ with $d_{Au}=51~$nm. The ratio between these two areas provides an estimation of the highest number 
of viral arms that can be grafted to the nanoparticle surface, $n_{max} \simeq 26$, very close to the average valency $n \simeq 23$ found experimentally (Fig.~\ref{fig:Star-components} and Supplementary Fig.~S1). 



\section*{Conclusion}

In this work, we have produced and studied the dynamics, in the dilute regime, of 
virus-based hybrid soft monodisperse star particles. Taking advantage of the optical contrast between the metallic Au core and the viral arms, we have shown thanks to optical microscopy and dynamic light scattering that, beyond their overall Brownian diffusion, these stars exhibit internal degrees of freedom which can stem both from the flexibility of the whole phage and from the virus tip proteins acting as entropic springs. A flexible link is evidenced between the 
virus body and the metallic core formed by the Au nanoparticle, whose dynamics can be accounted for by a spring model, \textit{i.e.} a Brownian particle trapped in a harmonic potential. The internal flexibility of the star particles combined with their sparse morphology make this system very appealing for probing the dense regimes and investigating their mutual interactions potentially leading to dynamical arrest in glassy or jammed states.    


\section*{\label{sec:experimental}Material and methods}

\subsection*{Hybrid virus-based star particles}  

The star particles are produced by self-assembly in solution of tannic-acid-stabilized Au (NPs) of  diameter $d_{Au} = 51 \pm 6$~nm (NanoComposix, Inc) with genetically modified rod-shaped viruses, named M13C7C. The M13C7C virus mutant, obtained from the commercially available phage display peptide library  
kit (New England Biolabs), is a \textit{monodisperse} filamentous bacteriophage of contour length $L_c=1 \ \mathrm{\mu m}$ and diameter $d = 7\ \mathrm{nm}$. 
This negatively charged polyelectrolyte is semi-rigid with a persistence length, $L_p$, exceeding its contour length, \textit{i.e.} $L_p \simeq 3 L_c$ \cite{Barry2009}. The 5 copies of the tip protein p3 displayed at the proximal end of the virus, are genetically modified with the fusion of the following amino-acid sequence before the N-terminal amine and therefore exposed to the solvent: GGG\underline{C}TAERVPD\underline{C}A-NH2 \cite{Repula2019}. 
The two cysteine residues, \underline{C}, present solely at one of the virus ends, and otherwise absent from the capsid, are thiol-containing amino acids, forming disulfide bonds 
which enable selective binding to noble metals \textit{via} a dative bond \cite{Hermanson-book}. 
Both components of the stars are initially mixed 
with the virus mutants in a molar excess of about 100 per Au NP. To promote the star formation by grafting the virus tips to the Au NPs, 
the electrostatic repulsion between Au NPs and viruses is screened step-wise by dialyzing the suspension against Tris/HCl/NaCl buffers at pH~8.2 in a range of ionic strength from I~=~20 mM to I~=~200 mM by adjusting the amount of NaCl. This results in star-like particles with viral arms linked by one tip to a solid metal core (Fig.~\ref{fig:Star-components}) \cite{delaCotte2017,Huang2009}. 
The excess of viruses is removed by typically three centrifugation steps at 4000~g for 30 min: the stars are collected in the pellet before being redispersed, whereas the supernatant containing the non-reacted viruses is discarded. After purification, hybrid virus-based star particles are obtained, as observed by transmission electron microscopy in Fig.~\ref{fig:Star-components} and Supplementary Fig.~S1, from which the distribution of arms, and therefore the average valency, $<n>$ = 23 $\pm$ 6 viruses per Au NP, is measured.
The fluorescent structures are prepared following the same protocol but with a batch of M13C7C previously labeled with red fluorescent dyes (Dylight 550-NHS ester, Thermo Scientific) according to a procedure reported elsewhere \cite{delaCotte2017}.  
Using the plasmonic resonance of the Au core 
at 
$\lambda_{\mathrm{Au}} = 520$~nm, the particle concentration of the hybrid virus-based stars is determined by $C_{\mathrm{star}} = C_{\mathrm{AuNP}} = \mathrm{OD_{520}} N_A / (\ell \times\varepsilon_{520})$, with $\mathrm{OD_{520}}$ the optical absorption, $N_A$ Avogadro's number, $\ell$ the optical path and $\varepsilon_{520} = 1.94 \times 10^{10} \ \mathrm{M^{-1} cm^{-1}}$ the Au NP molar extinction coefficient 
\cite{ExtCoeff}.

\subsection*{Transmission Electron Microscopy (TEM)}

Suspensions with typical concentrations $C_{\mathrm{star}} = \mathrm{1.6 \times 10^{9}}$ stars/mL are used to prepare samples for TEM on $\mathrm{O_2}$  plasma-treated carbon-coated grids then negatively stained with 2$\%$ uranyl acetate to reveal the organic viruses. An AMT CCD camera mounted on a Hitachi H-600 electron microscope operating at 75~kV is used to observe the grids and record the images. 


\subsection*{Optical microscopy and single particle tracking (SPT)}

Suspensions of stars of respective optical density $\mathrm{OD_{520}} = 0.05$ (\textit{i.e.} $C_{\mathrm{star}} = 1.6\times 10^{9}$ stars/mL) for fluorescence and confocal microscopy, and $\mathrm{OD_{520}} = 0.03$ (\textit{i.e.} $C_{\mathrm{star}} = 9\times 10^{8}$ stars/mL) for dark-field microscopy, are prepared in Tris/HCl/NaCl buffer (I~=~100~mM, pH~8.2) and set in 
optical cells whose thickness is between 10 and 15$~\mu$m. The microscopy cells are composed by a cover slip and a glass slide, first cleaned with sulfochromic acid, and sealed by ultraviolet-cured glue (NOA81, Epotecny).
Confocal fluorescence microscopy images are recorded with a Zeiss LSM 980 microscope at a frame rate of 35~fps.
The dark-field (DF) 
observations are performed using an inverted microscope equipped with an Olympus UPlanFLN 100x oil-immersion Objective of adjustable numerical aperture (NA) from 0.6 to 1.3, and an Olympus U-DCW oil-immersion dark-field condenser (NA=1.2-1.4). The same microscope, equipped with a Omicron LedHub as a light source, is used for fluorescence observation, where the objective is a 100x oil-immersion PlanAPO of high numerical aperture, NA= 1.4. Images are acquired with an ultra-fast electron-multiplying camera (NEO sCMOS Andor) having a pixel size of 6.5~$\mu$m. The dynamics of the system is obtained using single particle tracking (SPT) in wide-field microscopy, where we measure the two-dimensional projection of the isotropic Brownian trajectories in the objective focal plane. 
\cite{Crocker1996,Shen2017,Novotny2019} The experimental conditions are optimized to get the highest frame rate possible with a signal-to-noise ratio sufficient for a good detection. Thanks to the high scattering signal of the Au NPs, a frame rate of 762~fps was applied for the dark-field microscopy. It decreased to 199~fps for fluorescence microscopy whose signal-to-noise ratio is intrinsically lower. Two-dimensional traces \textbf{r}$(t)$ are collected using a custom-written particle tracking algorithm developed in MATLAB (MathWorks) (Supplementary Fig.~S2). The mean squared displacement defined as $\mathrm{MSD} = \langle \mathbf{r}^2(t) \rangle$, is calculated for each trace (Supplementary Fig.~S3), before being first averaged over the total number of detected particles 
and then fitted to determine the corresponding diffusion coefficient according to Eq.~\ref{MSD-eqfit} with $n=2$ \cite{Shen2017}. The number of valid collected traces acquired for each contrast mode is $\mathrm{N_{star}}=160$ and 151 for the dark-field and fluorescence microscopy techniques, respectively (Supplementary Fig.~S3). This high number of traces, $\mathrm{N_{star}} > 100$, allows for an accurate estimation of the diffusion coefficients with an error bar of less than 5$\%$ \cite{Novotny2019}. We cutoff our fitting range (Fig.~\ref{fig:MSD} and Supplementary Fig.~S3) because of the limitation on tracking particles for long times (larger than 1~s). 

\subsection*{Dynamic Light Scattering (DLS)}

Three samples having star concentration of $\mathrm{OD_{520}} = 0.15$ (\textit{i.e.} $C_{\mathrm{star}} = 4.7\times 10^{9}$ stars/mL) in I~=~0.5~mM pH~8.2 Tris-HCl buffer, as well as $\mathrm{OD_{520}} = 0.05$ (\textit{i.e.} $C_{\mathrm{star}} = 1.6\times 10^{9}$ stars/mL) and $\mathrm{OD_{520}} = 0.17$ (\textit{i.e.} $C_{\mathrm{star}} = 5.3\times 10^{9}$ stars/mL) in I~=~100~mM pH~8.2 Tris-HCl-NaCl buffer, are prepared by dialysis and then transferred  into a light scattering glass tube of 10~mm diameter. The tube is immersed in a bath of toluene for index matching, and regulated at a constant temperature of $T=20^\mathrm{o}$C.  
The time-intensity correlation function, $g_2(q,t)$, of the scattered light is recorded in the polarized (\textit{i.e.}, VV) geometry \cite{Pecora} using an ALV laser goniometer with a HeNe linearly polarized laser of $\mathrm{\lambda_L = 632.8}$~nm and an ALV500/EPP digital correlator (ALV-GmbH Technology). 
The scattering angle, $\theta$, is varied from $30^\mathrm{o}$ to $150^\mathrm{o}$ in steps of $10^\mathrm{o}$, which allows us to probe a range of scattering wavevectors

\begin{equation}
\label{Scattering Angle}
q = \frac{4 \pi n_0}{\mathrm{\lambda_L}}\sin(\theta /2) \mathrm{,}
\end{equation}

\noindent from  $\mathrm{6.8 \times 10^{-3} \ nm^{-1}} \leq q \leq \mathrm{2.55 \times 10^{-2} \ nm^{-1}}$, with $n_0$ the refractive index of the solvent -- here, water -- \cite{Pecora}. 
The thirteen scattered intensity correlation functions $g_2(q,t)$ covering the full range of scattering angles 
are then globally (\textit{i.e.} \textit{not} independently) fitted with Eq.~\ref{g2-spheres} using OriginPro~2020 (Originlab Corporation) software, accounting for the basic Brownian diffusion of spherical particles, or with Eq.~\ref{fit-spring}.
The latter 
is derived from the decoupling approximation of the intermediate scattering function in which $F_s(q,t)$ includes contributions from the overall translational (trans) and rotational (rot) diffusions as well as from internal dynamics of the particle, which can be modeled by a constrained Brownian motion within a harmonic potential as first described  by Ornstein \& Uhlenbeck \cite{Onrstein-Uhlenbeck,Fragments-link}

\begin{equation}
\label{Decopling FS}
F_s(q,t) = F_{trans}(q,t) \cdot F_{rot}(q,t) \cdot F_{int} (q,t) \mathrm{,}
\end{equation}

\noindent with $F_{trans}(q,t)$ and $F_{rot}(q,t)$ the translational and rotational contributions to the intermediate scattering function respectively, and 
		
\begin{equation}
\label{flex}
F_{int} = \exp  \left\lbrace -q^2\left\langle r^2_{int} \right\rangle   \left[1 - \exp(-\xi t) \right] \right\rbrace \mathrm{,}
\end{equation} 

\noindent is the 
contribution due to the internal dynamics (spring model) 
\cite{Fragments-link}. Different parameters characterize these internal degrees of freedom:
$\left\langle r^2_{int} \right\rangle = \frac{k_B T}{\kappa}$ with $k_B$ the Boltzmann constant and $T$ the temperature, is the mean square displacement within the harmonic potential at the relaxation time $1/\xi = \gamma_0/ \kappa$, written in terms of the friction coefficient, $\gamma_0$, and spring constant, $\kappa$. Considering that the NPs being probed are nearly spherical, we approximate the first two factors of Eq.~\ref{Decopling FS} to the intermediate scattering function of monodisperse isotropic particles \cite{Pecora}

\begin{equation}
\label{aprox-sphereFs}
     F_{trans}(q,t) \cdot F_{rot}(q,t) \approx F_{\mathrm{sphere}} (q,t) = \exp{(-q^2 D_T t)} \mathrm{.}
\end{equation}

\noindent Combining Eqs.~\ref{Decopling FS}, \ref{flex} and \ref{aprox-sphereFs}, 
and applying the relation between $F_s(q,t)$ and $g_2(q,t)$ given by Eq.~\ref{g2-spheres}, we obtain the final expression of Eq.~\ref{fit-spring}, which is used to globally fit the experimental data, as shown in Fig.~
\ref{fig:DLS-CS} by black lines. 

\bigbreak

\textbf{Supporting Information Available}

Electron microscopy images and valency of star particles;  traces obtained from single particle tracking and corresponding MSD of stars; DLS data and fitting with uniform sphere model and with the spring model at different dilutions; model of the random displacement of the apparent center-of-mass (PDF). Confocal and fluorescence microscopy movies of star particles (AVI).

\bigbreak



\textbf{Author Contributions}

A.B.Z.-M. prepared the samples and the first version of the manuscript, performed the experiments, collected and analyzed the data. E.G. devised the study, supervised the project, analyzed and interpreted the data, wrote, revised and edited the manuscript.  

\textbf{Notes}

The authors declare no competing financial interest.

\bigbreak

\textbf{Acknowledgments} 
We thank F.~Nallet for his interest in this work. We acknowledge financial support from the French National Research Agency (ANR) under grant No.~ANR-21-CE06-0045-StarRAC. 

\bibliography{CS}

\providecommand{\latin}[1]{#1}
\makeatletter
\providecommand{\doi}
  {\begingroup\let\do\@makeother\dospecials
  \catcode`\{=1 \catcode`\}=2\doi@aux}
\providecommand{\doi@aux}[1]{\endgroup\texttt{#1}}
\makeatother
\providecommand*\mcitethebibliography{\thebibliography}
\csname @ifundefined\endcsname{endmcitethebibliography}
  {\let\endmcitethebibliography\endthebibliography}{}
\begin{mcitethebibliography}{65}
\providecommand*\natexlab[1]{#1}
\providecommand*\mciteSetBstSublistMode[1]{}
\providecommand*\mciteSetBstMaxWidthForm[2]{}
\providecommand*\mciteBstWouldAddEndPuncttrue
  {\def\EndOfBibitem{\unskip.}}
\providecommand*\mciteBstWouldAddEndPunctfalse
  {\let\EndOfBibitem\relax}
\providecommand*\mciteSetBstMidEndSepPunct[3]{}
\providecommand*\mciteSetBstSublistLabelBeginEnd[3]{}
\providecommand*\EndOfBibitem{}
\mciteSetBstSublistMode{f}
\mciteSetBstMaxWidthForm{subitem}{(\alph{mcitesubitemcount})}
\mciteSetBstSublistLabelBeginEnd
  {\mcitemaxwidthsubitemform\space}
  {\relax}
  {\relax}

\bibitem[Vincent \latin{et~al.}(1986)Vincent, Edwards, Emmett, and
  Jones]{vincent_depletion_1986}
Vincent,~B.; Edwards,~J.; Emmett,~S.; Jones,~A. \emph{Colloids Surf.}
  \textbf{1986}, \emph{18}, 261--281\relax
\mciteBstWouldAddEndPuncttrue
\mciteSetBstMidEndSepPunct{\mcitedefaultmidpunct}
{\mcitedefaultendpunct}{\mcitedefaultseppunct}\relax
\EndOfBibitem
\bibitem[Likos(2006)]{Likos2006}
Likos,~C.~N. \emph{Soft Matter} \textbf{2006}, \emph{2}, 478--498\relax
\mciteBstWouldAddEndPuncttrue
\mciteSetBstMidEndSepPunct{\mcitedefaultmidpunct}
{\mcitedefaultendpunct}{\mcitedefaultseppunct}\relax
\EndOfBibitem
\bibitem[Vlassopoulos and Cloitre(2014)Vlassopoulos, and
  Cloitre]{VLASSOPOULOS2014561}
Vlassopoulos,~D.; Cloitre,~M. \emph{Curr. Opin. Colloid Interface Sci.}
  \textbf{2014}, \emph{19}, 561--574\relax
\mciteBstWouldAddEndPuncttrue
\mciteSetBstMidEndSepPunct{\mcitedefaultmidpunct}
{\mcitedefaultendpunct}{\mcitedefaultseppunct}\relax
\EndOfBibitem
\bibitem[Wei \latin{et~al.}(2013)Wei, Zhao, Hollingsworth, Zhou, Jin, Zhang,
  Cheng, and Han]{wei_mechanism_2013}
Wei,~G.; Zhao,~C.; Hollingsworth,~J.; Zhou,~Z.; Jin,~F.; Zhang,~Z.; Cheng,~H.;
  Han,~C.~C. \emph{Soft Matter} \textbf{2013}, \emph{9}, 9924\relax
\mciteBstWouldAddEndPuncttrue
\mciteSetBstMidEndSepPunct{\mcitedefaultmidpunct}
{\mcitedefaultendpunct}{\mcitedefaultseppunct}\relax
\EndOfBibitem
\bibitem[Lindenblatt \latin{et~al.}(2000)Lindenblatt, Schärtl, Pakula, and
  Schmidt]{lindenblatt_synthesis_2000}
Lindenblatt,~G.; Schärtl,~W.; Pakula,~T.; Schmidt,~M. \emph{Macromolecules}
  \textbf{2000}, \emph{33}, 9340--9347\relax
\mciteBstWouldAddEndPuncttrue
\mciteSetBstMidEndSepPunct{\mcitedefaultmidpunct}
{\mcitedefaultendpunct}{\mcitedefaultseppunct}\relax
\EndOfBibitem
\bibitem[Eckert and Richtering(2008)Eckert, and Richtering]{Eckert2008}
Eckert,~T.; Richtering,~W. \emph{J. Chem. Phys.} \textbf{2008}, \emph{129},
  124902\relax
\mciteBstWouldAddEndPuncttrue
\mciteSetBstMidEndSepPunct{\mcitedefaultmidpunct}
{\mcitedefaultendpunct}{\mcitedefaultseppunct}\relax
\EndOfBibitem
\bibitem[Laurati \latin{et~al.}(2005)Laurati, Stellbrink, Lund, Willner,
  Richter, and Zaccarelli]{laurati_starlike_2005}
Laurati,~M.; Stellbrink,~J.; Lund,~R.; Willner,~L.; Richter,~D.; Zaccarelli,~E.
  \emph{Phys. Rev. Lett.} \textbf{2005}, \emph{94}, 195504\relax
\mciteBstWouldAddEndPuncttrue
\mciteSetBstMidEndSepPunct{\mcitedefaultmidpunct}
{\mcitedefaultendpunct}{\mcitedefaultseppunct}\relax
\EndOfBibitem
\bibitem[Merlet-Lacroix \latin{et~al.}(2010)Merlet-Lacroix, Di~Cola, and
  Cloitre]{merlet-lacroix_swelling_2010}
Merlet-Lacroix,~N.; Di~Cola,~E.; Cloitre,~M. \emph{Soft Matter} \textbf{2010},
  \emph{6}, 984\relax
\mciteBstWouldAddEndPuncttrue
\mciteSetBstMidEndSepPunct{\mcitedefaultmidpunct}
{\mcitedefaultendpunct}{\mcitedefaultseppunct}\relax
\EndOfBibitem
\bibitem[Erwin \latin{et~al.}(2010)Erwin, Cloitre, Gauthier, and
  Vlassopoulos]{erwin_dynamics_2010}
Erwin,~B.~M.; Cloitre,~M.; Gauthier,~M.; Vlassopoulos,~D. \emph{Soft Matter}
  \textbf{2010}, \emph{6}, 2825--2833\relax
\mciteBstWouldAddEndPuncttrue
\mciteSetBstMidEndSepPunct{\mcitedefaultmidpunct}
{\mcitedefaultendpunct}{\mcitedefaultseppunct}\relax
\EndOfBibitem
\bibitem[Gupta \latin{et~al.}(2015)Gupta, Camargo, Stellbrink, Allgaier,
  Radulescu, Lindner, Zaccarelli, Likos, and Richter]{Gupta2015}
Gupta,~S.; Camargo,~M.; Stellbrink,~J.; Allgaier,~J.; Radulescu,~A.;
  Lindner,~P.; Zaccarelli,~E.; Likos,~C.~N.; Richter,~D. \emph{Nanoscale}
  \textbf{2015}, \emph{7}, 13924--13934\relax
\mciteBstWouldAddEndPuncttrue
\mciteSetBstMidEndSepPunct{\mcitedefaultmidpunct}
{\mcitedefaultendpunct}{\mcitedefaultseppunct}\relax
\EndOfBibitem
\bibitem[Gury \latin{et~al.}(2019)Gury, Gauthier, Cloitre, and
  Vlassopoulos]{gury2019colloidal}
Gury,~L.; Gauthier,~M.; Cloitre,~M.; Vlassopoulos,~D. \emph{Macromolecules}
  \textbf{2019}, \emph{52}, 4617--4623\relax
\mciteBstWouldAddEndPuncttrue
\mciteSetBstMidEndSepPunct{\mcitedefaultmidpunct}
{\mcitedefaultendpunct}{\mcitedefaultseppunct}\relax
\EndOfBibitem
\bibitem[Pusey and van Megen(1986)Pusey, and van Megen]{pusey_phase_1986}
Pusey,~P.~N.; van Megen,~W. \emph{Nature} \textbf{1986}, \emph{320},
  340--342\relax
\mciteBstWouldAddEndPuncttrue
\mciteSetBstMidEndSepPunct{\mcitedefaultmidpunct}
{\mcitedefaultendpunct}{\mcitedefaultseppunct}\relax
\EndOfBibitem
\bibitem[Tokuyama and Oppenheim(1994)Tokuyama, and
  Oppenheim]{tokuyama1994dynamics}
Tokuyama,~M.; Oppenheim,~I. \emph{Phys. Rev. E} \textbf{1994}, \emph{50},
  R16\relax
\mciteBstWouldAddEndPuncttrue
\mciteSetBstMidEndSepPunct{\mcitedefaultmidpunct}
{\mcitedefaultendpunct}{\mcitedefaultseppunct}\relax
\EndOfBibitem
\bibitem[Pusey and Tough(1983)Pusey, and Tough]{pusey1983hydrodynamic}
Pusey,~P.~N.; Tough,~R.~J. \emph{Faraday Discuss. Chem. Soc.} \textbf{1983},
  \emph{76}, 123--136\relax
\mciteBstWouldAddEndPuncttrue
\mciteSetBstMidEndSepPunct{\mcitedefaultmidpunct}
{\mcitedefaultendpunct}{\mcitedefaultseppunct}\relax
\EndOfBibitem
\bibitem[van Megen \latin{et~al.}(1991)van Megen, Underwood, and
  Pusey]{van1991dynamics}
van Megen,~W.; Underwood,~S.~M.; Pusey,~P.~N. \emph{J. Chem. Soc., Faraday
  Trans.} \textbf{1991}, \emph{87}, 395--401\relax
\mciteBstWouldAddEndPuncttrue
\mciteSetBstMidEndSepPunct{\mcitedefaultmidpunct}
{\mcitedefaultendpunct}{\mcitedefaultseppunct}\relax
\EndOfBibitem
\bibitem[Vlassopoulos and Cloitre(2021)Vlassopoulos, and
  Cloitre]{vlassopoulos2021suspensions}
Vlassopoulos,~D.; Cloitre,~M. \emph{Theory and Applications of Colloidal
  Suspension Rheology}; Cambridge University Press Cambridge, 2021; pp
  227--290\relax
\mciteBstWouldAddEndPuncttrue
\mciteSetBstMidEndSepPunct{\mcitedefaultmidpunct}
{\mcitedefaultendpunct}{\mcitedefaultseppunct}\relax
\EndOfBibitem
\bibitem[De~Gennes and Leger(1982)De~Gennes, and Leger]{de1982dynamics}
De~Gennes,~P.; Leger,~L. \emph{Annu. Rev. Phys. Chem.} \textbf{1982},
  \emph{33}, 49--61\relax
\mciteBstWouldAddEndPuncttrue
\mciteSetBstMidEndSepPunct{\mcitedefaultmidpunct}
{\mcitedefaultendpunct}{\mcitedefaultseppunct}\relax
\EndOfBibitem
\bibitem[Chu \latin{et~al.}(1991)Chu, Wang, and Yu]{chu1991dynamic}
Chu,~B.; Wang,~Z.; Yu,~J. \emph{Macromolecules} \textbf{1991}, \emph{24},
  6832--6838\relax
\mciteBstWouldAddEndPuncttrue
\mciteSetBstMidEndSepPunct{\mcitedefaultmidpunct}
{\mcitedefaultendpunct}{\mcitedefaultseppunct}\relax
\EndOfBibitem
\bibitem[Romeo \latin{et~al.}(2012)Romeo, Imperiali, Kim, Fern{\'a}ndez-Nieves,
  and Weitz]{romeo2012origin}
Romeo,~G.; Imperiali,~L.; Kim,~J.-W.; Fern{\'a}ndez-Nieves,~A.; Weitz,~D.~A.
  \emph{J. Chem. Phys.} \textbf{2012}, \emph{136}, 124905\relax
\mciteBstWouldAddEndPuncttrue
\mciteSetBstMidEndSepPunct{\mcitedefaultmidpunct}
{\mcitedefaultendpunct}{\mcitedefaultseppunct}\relax
\EndOfBibitem
\bibitem[Freedman \latin{et~al.}(2005)Freedman, Lee, Li, Luo, Skobeleva, and
  Ke]{freedman2005diffusion}
Freedman,~K.~O.; Lee,~J.; Li,~Y.; Luo,~D.; Skobeleva,~V.~B.; Ke,~P.~C. \emph{J.
  Phys. Chem. B} \textbf{2005}, \emph{109}, 9839--9842\relax
\mciteBstWouldAddEndPuncttrue
\mciteSetBstMidEndSepPunct{\mcitedefaultmidpunct}
{\mcitedefaultendpunct}{\mcitedefaultseppunct}\relax
\EndOfBibitem
\bibitem[Cautela \latin{et~al.}(2020)Cautela, Stenqvist, Schill{\'e}n,
  Beli{\'{c}}, M{\aa}nsson, Hagemans, Seuss, Fery, Crassous, and
  Galantini]{Cautela2020}
Cautela,~J.; Stenqvist,~B.; Schill{\'e}n,~K.; Beli{\'{c}},~D.;
  M{\aa}nsson,~L.~K.; Hagemans,~F.; Seuss,~M.; Fery,~A.; Crassous,~J.~J.;
  Galantini,~L. \emph{ACS Nano} \textbf{2020}, \emph{14}, 15748--15756\relax
\mciteBstWouldAddEndPuncttrue
\mciteSetBstMidEndSepPunct{\mcitedefaultmidpunct}
{\mcitedefaultendpunct}{\mcitedefaultseppunct}\relax
\EndOfBibitem
\bibitem[Gasser \latin{et~al.}(2014)Gasser, Hyatt, Lietor-Santos, Herman, Lyon,
  and Fernandez-Nieves]{gasser_form_2014}
Gasser,~U.; Hyatt,~J.~S.; Lietor-Santos,~J.-J.; Herman,~E.~S.; Lyon,~L.~A.;
  Fernandez-Nieves,~A. \emph{J. Chem. Phys.} \textbf{2014}, \emph{141},
  034901\relax
\mciteBstWouldAddEndPuncttrue
\mciteSetBstMidEndSepPunct{\mcitedefaultmidpunct}
{\mcitedefaultendpunct}{\mcitedefaultseppunct}\relax
\EndOfBibitem
\bibitem[Laurati \latin{et~al.}(2005)Laurati, Stellbrink, Lund, Willner,
  Richter, and Zaccarelli]{laurati2005starlike}
Laurati,~M.; Stellbrink,~J.; Lund,~R.; Willner,~L.; Richter,~D.; Zaccarelli,~E.
  \emph{Phys. Rev. Lett.} \textbf{2005}, \emph{94}, 195504\relax
\mciteBstWouldAddEndPuncttrue
\mciteSetBstMidEndSepPunct{\mcitedefaultmidpunct}
{\mcitedefaultendpunct}{\mcitedefaultseppunct}\relax
\EndOfBibitem
\bibitem[Fleischer \latin{et~al.}(2000)Fleischer, Fytas, Vlassopoulos, Roovers,
  and Hadjichristidis]{fleischer_self-di_2000}
Fleischer,~G.; Fytas,~G.; Vlassopoulos,~D.; Roovers,~J.; Hadjichristidis,~N.
  \emph{Phys. A (Amsterdam, Neth.)} \textbf{2000}, \emph{280}, 266--278\relax
\mciteBstWouldAddEndPuncttrue
\mciteSetBstMidEndSepPunct{\mcitedefaultmidpunct}
{\mcitedefaultendpunct}{\mcitedefaultseppunct}\relax
\EndOfBibitem
\bibitem[Mattsson \latin{et~al.}(2009)Mattsson, Wyss, Fernandez-Nieves,
  Miyazaki, Hu, Reichman, and Weitz]{Mattsson2009}
Mattsson,~J.; Wyss,~H.~M.; Fernandez-Nieves,~A.; Miyazaki,~K.; Hu,~Z.;
  Reichman,~D.~R.; Weitz,~D.~A. \emph{Nature} \textbf{2009}, \emph{462},
  83--86\relax
\mciteBstWouldAddEndPuncttrue
\mciteSetBstMidEndSepPunct{\mcitedefaultmidpunct}
{\mcitedefaultendpunct}{\mcitedefaultseppunct}\relax
\EndOfBibitem
\bibitem[Conley \latin{et~al.}(2017)Conley, Aebischer, Nöjd, Schurtenberger,
  and Scheffold]{Conley2017}
Conley,~G.~M.; Aebischer,~P.; Nöjd,~S.; Schurtenberger,~P.; Scheffold,~F.
  \emph{Sci. Adv.} \textbf{2017}, \emph{3}, e1700969\relax
\mciteBstWouldAddEndPuncttrue
\mciteSetBstMidEndSepPunct{\mcitedefaultmidpunct}
{\mcitedefaultendpunct}{\mcitedefaultseppunct}\relax
\EndOfBibitem
\bibitem[Zhang \latin{et~al.}(2014)Zhang, Lettinga, Dhont, and
  Stiakakis]{Manolis2014}
Zhang,~J.; Lettinga,~P.~M.; Dhont,~J. K.~G.; Stiakakis,~E. \emph{Phys. Rev.
  Lett.} \textbf{2014}, \emph{113}, 268303\relax
\mciteBstWouldAddEndPuncttrue
\mciteSetBstMidEndSepPunct{\mcitedefaultmidpunct}
{\mcitedefaultendpunct}{\mcitedefaultseppunct}\relax
\EndOfBibitem
\bibitem[Romero-Sanchez \latin{et~al.}(2022)Romero-Sanchez, Pihlajamaa,
  Adžić, Castellano, Stiakakis, Likos, and Laurati]{Manolis2022}
Romero-Sanchez,~I.; Pihlajamaa,~I.; Adžić,~N.; Castellano,~L.~E.;
  Stiakakis,~E.; Likos,~C.~N.; Laurati,~M. \emph{ACS Nano} \textbf{2022},
  \emph{16}, 2133--2146\relax
\mciteBstWouldAddEndPuncttrue
\mciteSetBstMidEndSepPunct{\mcitedefaultmidpunct}
{\mcitedefaultendpunct}{\mcitedefaultseppunct}\relax
\EndOfBibitem
\bibitem[Buitenhuis and Förster(1997)Buitenhuis, and
  Förster]{buitenhuis_block_1997}
Buitenhuis,~J.; Förster,~S. \emph{J. Chem. Phys.} \textbf{1997}, \emph{107},
  262--272\relax
\mciteBstWouldAddEndPuncttrue
\mciteSetBstMidEndSepPunct{\mcitedefaultmidpunct}
{\mcitedefaultendpunct}{\mcitedefaultseppunct}\relax
\EndOfBibitem
\bibitem[Vlassopoulos \latin{et~al.}(2001)Vlassopoulos, Fytas, Pakula, and
  Roovers]{vlassopoulos_multiarm_2001}
Vlassopoulos,~D.; Fytas,~G.; Pakula,~T.; Roovers,~J. \emph{J. Phys.: Condens.
  Matter} \textbf{2001}, \emph{13}, R855--R876\relax
\mciteBstWouldAddEndPuncttrue
\mciteSetBstMidEndSepPunct{\mcitedefaultmidpunct}
{\mcitedefaultendpunct}{\mcitedefaultseppunct}\relax
\EndOfBibitem
\bibitem[Cloitre(2010)]{cloitre_high_2010}
Cloitre,~M., Ed. \emph{High Solid Dispersions}; Advances in Polymer Science;
  Springer Berlin Heidelberg, 2010; Vol. 236\relax
\mciteBstWouldAddEndPuncttrue
\mciteSetBstMidEndSepPunct{\mcitedefaultmidpunct}
{\mcitedefaultendpunct}{\mcitedefaultseppunct}\relax
\EndOfBibitem
\bibitem[Choi \latin{et~al.}(2012)Choi, Ming~Hui, Pietrasik, Dong,
  Matyjaszewski, and R.~Bockstaller]{choi_toughening_2012}
Choi,~J.; Ming~Hui,~C.; Pietrasik,~J.; Dong,~H.; Matyjaszewski,~K.;
  R.~Bockstaller,~M. \emph{Soft Matter} \textbf{2012}, \emph{8},
  4072--4082\relax
\mciteBstWouldAddEndPuncttrue
\mciteSetBstMidEndSepPunct{\mcitedefaultmidpunct}
{\mcitedefaultendpunct}{\mcitedefaultseppunct}\relax
\EndOfBibitem
\bibitem[Shay \latin{et~al.}(2001)Shay, Raghavan, and
  Khan]{shay2001thermoreversible}
Shay,~J.~S.; Raghavan,~S.~R.; Khan,~S.~A. \emph{J. Rheol.} \textbf{2001},
  \emph{45}, 913--927\relax
\mciteBstWouldAddEndPuncttrue
\mciteSetBstMidEndSepPunct{\mcitedefaultmidpunct}
{\mcitedefaultendpunct}{\mcitedefaultseppunct}\relax
\EndOfBibitem
\bibitem[Ohno \latin{et~al.}(2006)Ohno, Morinaga, Takeno, Tsujii, and
  Fukuda]{ohno2006suspensions}
Ohno,~K.; Morinaga,~T.; Takeno,~S.; Tsujii,~Y.; Fukuda,~T.
  \emph{Macromolecules} \textbf{2006}, \emph{39}, 1245--1249\relax
\mciteBstWouldAddEndPuncttrue
\mciteSetBstMidEndSepPunct{\mcitedefaultmidpunct}
{\mcitedefaultendpunct}{\mcitedefaultseppunct}\relax
\EndOfBibitem
\bibitem[Ohno \latin{et~al.}(2007)Ohno, Morinaga, Takeno, Tsujii, and
  Fukuda]{ohno_suspensions_2007}
Ohno,~K.; Morinaga,~T.; Takeno,~S.; Tsujii,~Y.; Fukuda,~T.
  \emph{Macromolecules} \textbf{2007}, \emph{40}, 9143--9150\relax
\mciteBstWouldAddEndPuncttrue
\mciteSetBstMidEndSepPunct{\mcitedefaultmidpunct}
{\mcitedefaultendpunct}{\mcitedefaultseppunct}\relax
\EndOfBibitem
\bibitem[Tricot(1986)]{tricot1986chain}
Tricot,~M. \emph{Macromolecules} \textbf{1986}, \emph{19}, 1268--1270\relax
\mciteBstWouldAddEndPuncttrue
\mciteSetBstMidEndSepPunct{\mcitedefaultmidpunct}
{\mcitedefaultendpunct}{\mcitedefaultseppunct}\relax
\EndOfBibitem
\bibitem[Lee \latin{et~al.}(2008)Lee, Venable, {MacKerell}, and
  Pastor]{lee_molecular_2008}
Lee,~H.; Venable,~R.~M.; {MacKerell},~A.~D.; Pastor,~R.~W. \emph{Biophys. J.}
  \textbf{2008}, \emph{95}, 1590--1599\relax
\mciteBstWouldAddEndPuncttrue
\mciteSetBstMidEndSepPunct{\mcitedefaultmidpunct}
{\mcitedefaultendpunct}{\mcitedefaultseppunct}\relax
\EndOfBibitem
\bibitem[Biffi \latin{et~al.}(2013)Biffi, Cerbino, Bomboi, Paraboschi, Asselta,
  Sciortino, and Bellini]{biffi_phase_2013}
Biffi,~S.; Cerbino,~R.; Bomboi,~F.; Paraboschi,~E.~M.; Asselta,~R.;
  Sciortino,~F.; Bellini,~T. \emph{Proc. Natl. Acad. Sci. U.S.A.}
  \textbf{2013}, \emph{110}, 15633--15637\relax
\mciteBstWouldAddEndPuncttrue
\mciteSetBstMidEndSepPunct{\mcitedefaultmidpunct}
{\mcitedefaultendpunct}{\mcitedefaultseppunct}\relax
\EndOfBibitem
\bibitem[Rovigatti \latin{et~al.}(2014)Rovigatti, Smallenburg, Romano, and
  Sciortino]{Rovigatti-Sciortino}
Rovigatti,~L.; Smallenburg,~F.; Romano,~F.; Sciortino,~F. \emph{ACS Nano}
  \textbf{2014}, \emph{8}, 3567--3574\relax
\mciteBstWouldAddEndPuncttrue
\mciteSetBstMidEndSepPunct{\mcitedefaultmidpunct}
{\mcitedefaultendpunct}{\mcitedefaultseppunct}\relax
\EndOfBibitem
\bibitem[Brady \latin{et~al.}(2017)Brady, Brooks, Cicuta, and
  Di~Michele]{brady_crystallization_2017}
Brady,~R.~A.; Brooks,~N.~J.; Cicuta,~P.; Di~Michele,~L. \emph{Nano Lett.}
  \textbf{2017}, \emph{17}, 3276--3281\relax
\mciteBstWouldAddEndPuncttrue
\mciteSetBstMidEndSepPunct{\mcitedefaultmidpunct}
{\mcitedefaultendpunct}{\mcitedefaultseppunct}\relax
\EndOfBibitem
\bibitem[Lattuada \latin{et~al.}(2022)Lattuada, Pietrangeli, and
  Sciortino]{Lattuada-Sciortino}
Lattuada,~E.; Pietrangeli,~T.; Sciortino,~F. \emph{J. Chem. Phys.}
  \textbf{2022}, \emph{157}, 135101\relax
\mciteBstWouldAddEndPuncttrue
\mciteSetBstMidEndSepPunct{\mcitedefaultmidpunct}
{\mcitedefaultendpunct}{\mcitedefaultseppunct}\relax
\EndOfBibitem
\bibitem[Lu \latin{et~al.}(2002)Lu, Weers, and Stellwagen]{lu_dna_2002}
Lu,~Y.; Weers,~B.; Stellwagen,~N.~C. \emph{Biopolymers} \textbf{2002},
  \emph{61}, 261--275\relax
\mciteBstWouldAddEndPuncttrue
\mciteSetBstMidEndSepPunct{\mcitedefaultmidpunct}
{\mcitedefaultendpunct}{\mcitedefaultseppunct}\relax
\EndOfBibitem
\bibitem[Brady \latin{et~al.}(2019)Brady, Kaufhold, Brooks, Foderà, and
  Di~Michele]{brady_flexibility_2019}
Brady,~R.~A.; Kaufhold,~W.~T.; Brooks,~N.~J.; Foderà,~V.; Di~Michele,~L.
  \emph{J. Phys.: Condens. Matter} \textbf{2019}, \emph{31}, 074003\relax
\mciteBstWouldAddEndPuncttrue
\mciteSetBstMidEndSepPunct{\mcitedefaultmidpunct}
{\mcitedefaultendpunct}{\mcitedefaultseppunct}\relax
\EndOfBibitem
\bibitem[de~la Cotte \latin{et~al.}(2017)de~la Cotte, Wu, Trevisan, Repula, and
  Grelet]{delaCotte2017}
de~la Cotte,~A.; Wu,~C.; Trevisan,~M.; Repula,~A.; Grelet,~E. \emph{ACS Nano}
  \textbf{2017}, \emph{11}, 10616--10622\relax
\mciteBstWouldAddEndPuncttrue
\mciteSetBstMidEndSepPunct{\mcitedefaultmidpunct}
{\mcitedefaultendpunct}{\mcitedefaultseppunct}\relax
\EndOfBibitem
\bibitem[Zhan \latin{et~al.}(2022)Zhan, Fang, Chen, Xiong, Guo, Huang, Li,
  Leng, Huang, and Xiong]{Zhan2022}
Zhan,~S.; Fang,~H.; Chen,~Q.; Xiong,~S.; Guo,~Y.; Huang,~T.; Li,~X.; Leng,~Y.;
  Huang,~X.; Xiong,~Y. \emph{Biosens. Bioelectron.} \textbf{2022}, \emph{217},
  114693\relax
\mciteBstWouldAddEndPuncttrue
\mciteSetBstMidEndSepPunct{\mcitedefaultmidpunct}
{\mcitedefaultendpunct}{\mcitedefaultseppunct}\relax
\EndOfBibitem
\bibitem[Mosquera \latin{et~al.}(2020)Mosquera, Garc{\'i}a, Henriksen-Lacey,
  Mart{\'i}nez-Calvo, Dhanjani, Mascare{\~{n}}as, and
  Liz-Marz{\'a}n]{Mosquera2020}
Mosquera,~J.; Garc{\'i}a,~I.; Henriksen-Lacey,~M.; Mart{\'i}nez-Calvo,~M.;
  Dhanjani,~M.; Mascare{\~{n}}as,~J.~L.; Liz-Marz{\'a}n,~L.~M. \emph{ACS Nano}
  \textbf{2020}, \emph{14}, 5382--5391\relax
\mciteBstWouldAddEndPuncttrue
\mciteSetBstMidEndSepPunct{\mcitedefaultmidpunct}
{\mcitedefaultendpunct}{\mcitedefaultseppunct}\relax
\EndOfBibitem
\bibitem[Song \latin{et~al.}(1991)Song, Kim, Wilcoxon, and
  Schurr]{song1991dynamic}
Song,~L.; Kim,~U.-S.; Wilcoxon,~J.; Schurr,~J.~M. \emph{Biopolymers}
  \textbf{1991}, \emph{31}, 547--567\relax
\mciteBstWouldAddEndPuncttrue
\mciteSetBstMidEndSepPunct{\mcitedefaultmidpunct}
{\mcitedefaultendpunct}{\mcitedefaultseppunct}\relax
\EndOfBibitem
\bibitem[Crocker and Grier(1996)Crocker, and Grier]{Crocker1996}
Crocker,~J.; Grier,~D.~G. \emph{J. Colloid Interface Sci.} \textbf{1996},
  \emph{179}, 298--310\relax
\mciteBstWouldAddEndPuncttrue
\mciteSetBstMidEndSepPunct{\mcitedefaultmidpunct}
{\mcitedefaultendpunct}{\mcitedefaultseppunct}\relax
\EndOfBibitem
\bibitem[Alvarez \latin{et~al.}(2017)Alvarez, Lettinga, and Grelet]{Eric-Laura}
Alvarez,~L.; Lettinga,~M.~P.; Grelet,~E. \emph{Phys. Rev. Lett.} \textbf{2017},
  \emph{118}, 178002\relax
\mciteBstWouldAddEndPuncttrue
\mciteSetBstMidEndSepPunct{\mcitedefaultmidpunct}
{\mcitedefaultendpunct}{\mcitedefaultseppunct}\relax
\EndOfBibitem
\bibitem[Stetefeld \latin{et~al.}(2016)Stetefeld, McKenna, and
  Patel]{Stetefeld2016}
Stetefeld,~J.; McKenna,~S.~A.; Patel,~T.~R. \emph{Biophys. Rev.} \textbf{2016},
  \emph{8}, 409--427\relax
\mciteBstWouldAddEndPuncttrue
\mciteSetBstMidEndSepPunct{\mcitedefaultmidpunct}
{\mcitedefaultendpunct}{\mcitedefaultseppunct}\relax
\EndOfBibitem
\bibitem[Berne and Pecora(2000)Berne, and Pecora]{Pecora}
Berne,~B.~J.; Pecora,~R. \emph{Dynamic light scattering. {W}ith applications to
  chemistry, biology, and physics}; Courier Corporation, 2000\relax
\mciteBstWouldAddEndPuncttrue
\mciteSetBstMidEndSepPunct{\mcitedefaultmidpunct}
{\mcitedefaultendpunct}{\mcitedefaultseppunct}\relax
\EndOfBibitem
\bibitem[Uhlenbeck and Ornstein(1930)Uhlenbeck, and
  Ornstein]{Onrstein-Uhlenbeck}
Uhlenbeck,~G.~E.; Ornstein,~L.~S. \emph{Phys. Rev.} \textbf{1930}, \emph{36},
  823--841\relax
\mciteBstWouldAddEndPuncttrue
\mciteSetBstMidEndSepPunct{\mcitedefaultmidpunct}
{\mcitedefaultendpunct}{\mcitedefaultseppunct}\relax
\EndOfBibitem
\bibitem[Stingaciu \latin{et~al.}(2016)Stingaciu, Ivanova, Ohl, Biehl, and
  Richter]{Fragments-link}
Stingaciu,~L.~R.; Ivanova,~O.; Ohl,~M.; Biehl,~R.; Richter,~D. \emph{Sci. Rep.}
  \textbf{2016}, \emph{6}, 1--13\relax
\mciteBstWouldAddEndPuncttrue
\mciteSetBstMidEndSepPunct{\mcitedefaultmidpunct}
{\mcitedefaultendpunct}{\mcitedefaultseppunct}\relax
\EndOfBibitem
\bibitem[Kratky and Porod(1949)Kratky, and Porod]{Kratky1949b}
Kratky,~O.; Porod,~G. \emph{Recl. Trav. Chim. Pays-Bas} \textbf{1949},
  \emph{68}, 1106--1122\relax
\mciteBstWouldAddEndPuncttrue
\mciteSetBstMidEndSepPunct{\mcitedefaultmidpunct}
{\mcitedefaultendpunct}{\mcitedefaultseppunct}\relax
\EndOfBibitem
\bibitem[Marvin \latin{et~al.}(2014)Marvin, Symmons, and Straus]{Marvin2014}
Marvin,~D.~A.; Symmons,~M.~F.; Straus,~S.~K. \emph{Prog. Biophys. Mol. Biol.}
  \textbf{2014}, \emph{114}, 80--122\relax
\mciteBstWouldAddEndPuncttrue
\mciteSetBstMidEndSepPunct{\mcitedefaultmidpunct}
{\mcitedefaultendpunct}{\mcitedefaultseppunct}\relax
\EndOfBibitem
\bibitem[Gray \latin{et~al.}(1981)Gray, Brown, and Marvin]{gray1981adsorption}
Gray,~C.~W.; Brown,~R.; Marvin,~D. \emph{J. Mol. Biol.} \textbf{1981},
  \emph{146}, 621--627\relax
\mciteBstWouldAddEndPuncttrue
\mciteSetBstMidEndSepPunct{\mcitedefaultmidpunct}
{\mcitedefaultendpunct}{\mcitedefaultseppunct}\relax
\EndOfBibitem
\bibitem[Conners \latin{et~al.}(2023)Conners, Le{\'o}n-Quezada, McLaren,
  Bennett, Daum, Rakonjac, and Gold]{Conners2023}
Conners,~R.; Le{\'o}n-Quezada,~R.~I.; McLaren,~M.; Bennett,~N.~J.; Daum,~B.;
  Rakonjac,~J.; Gold,~V. A.~M. \emph{Nat. Commun.} \textbf{2023}, \emph{14},
  2724\relax
\mciteBstWouldAddEndPuncttrue
\mciteSetBstMidEndSepPunct{\mcitedefaultmidpunct}
{\mcitedefaultendpunct}{\mcitedefaultseppunct}\relax
\EndOfBibitem
\bibitem[Barry and Dogic(2009)Barry, and Dogic]{Barry2009}
Barry,~E.; Dogic,~Z. \emph{Soft Matter} \textbf{2009}, \emph{5},
  2563--2570\relax
\mciteBstWouldAddEndPuncttrue
\mciteSetBstMidEndSepPunct{\mcitedefaultmidpunct}
{\mcitedefaultendpunct}{\mcitedefaultseppunct}\relax
\EndOfBibitem
\bibitem[Repula \latin{et~al.}(2019)Repula, Oshima~Menegon, Wu, van~der Schoot,
  and Grelet]{Repula2019}
Repula,~A.; Oshima~Menegon,~M.; Wu,~C.; van~der Schoot,~P.; Grelet,~E.
  \emph{Phys. Rev. Lett.} \textbf{2019}, \emph{122}, 128008\relax
\mciteBstWouldAddEndPuncttrue
\mciteSetBstMidEndSepPunct{\mcitedefaultmidpunct}
{\mcitedefaultendpunct}{\mcitedefaultseppunct}\relax
\EndOfBibitem
\bibitem[Hermanson(2008)]{Hermanson-book}
Hermanson,~G.~T. \emph{Bioconjugate Techniques}, 2nd ed.; Academic Press,
  2008\relax
\mciteBstWouldAddEndPuncttrue
\mciteSetBstMidEndSepPunct{\mcitedefaultmidpunct}
{\mcitedefaultendpunct}{\mcitedefaultseppunct}\relax
\EndOfBibitem
\bibitem[Huang \latin{et~al.}(2009)Huang, Addas, Ward, Flynn, Velasco, Hagan,
  Dogic, and Fraden]{Huang2009}
Huang,~F.; Addas,~K.; Ward,~A.; Flynn,~N.~T.; Velasco,~E.; Hagan,~M.~F.;
  Dogic,~Z.; Fraden,~S. \emph{Phys. Rev. Lett.} \textbf{2009}, \emph{102},
  108302\relax
\mciteBstWouldAddEndPuncttrue
\mciteSetBstMidEndSepPunct{\mcitedefaultmidpunct}
{\mcitedefaultendpunct}{\mcitedefaultseppunct}\relax
\EndOfBibitem
\bibitem[Liu \latin{et~al.}(2007)Liu, Atwater, Wang, and Huo]{ExtCoeff}
Liu,~X.; Atwater,~M.; Wang,~J.; Huo,~Q. \emph{Colloids Surf. B} \textbf{2007},
  \emph{58}, 3--7\relax
\mciteBstWouldAddEndPuncttrue
\mciteSetBstMidEndSepPunct{\mcitedefaultmidpunct}
{\mcitedefaultendpunct}{\mcitedefaultseppunct}\relax
\EndOfBibitem
\bibitem[Shen \latin{et~al.}(2017)Shen, Tauzin, Baiyasi, Wang, Moringo, Shuang,
  and Landes]{Shen2017}
Shen,~H.; Tauzin,~L.~J.; Baiyasi,~R.; Wang,~W.; Moringo,~N.; Shuang,~B.;
  Landes,~C.~F. \emph{Chem. Rev.} \textbf{2017}, \emph{117}, 7331--7376\relax
\mciteBstWouldAddEndPuncttrue
\mciteSetBstMidEndSepPunct{\mcitedefaultmidpunct}
{\mcitedefaultendpunct}{\mcitedefaultseppunct}\relax
\EndOfBibitem
\bibitem[Novotn{\'y} and Pumera(2019)Novotn{\'y}, and Pumera]{Novotny2019}
Novotn{\'y},~F.; Pumera,~M. \emph{Sci. Rep.} \textbf{2019}, \emph{9},
  13222\relax
\mciteBstWouldAddEndPuncttrue
\mciteSetBstMidEndSepPunct{\mcitedefaultmidpunct}
{\mcitedefaultendpunct}{\mcitedefaultseppunct}\relax
\EndOfBibitem
\end{mcitethebibliography}

\end{document}